Hindawi

*Research Article*

# Scale-Free Networks with the Same Degree Distribution: Different Structural Properties

**José H. H. Grisi-Filho, Raul Ossada, Fernando Ferreira, and Marcos Amaku**

*Faculdade de Medicina Veterinária e Zootecnia, Universidade de São Paulo, 05508-270 São Paulo, SP, Brazil*

Correspondence should be addressed to José H. H. Grisi-Filho; grisi@vps.fmvz.usp.br

We have analysed some structural properties of scale-free networks with the same degree distribution. Departing from a degree distribution obtained from the Barabási-Albert (BA) algorithm, networks were generated using four additional different algorithms (Molloy-Reed, Kalisky, and two new models named A and B) besides the BA algorithm itself. For each network, we have calculated the following structural measures: average degree of the nearest neighbours, central point dominance, clustering coefficient, the Pearson correlation coefficient, and global efficiency. We found that different networks with the same degree distribution may have distinct structural properties. In particular, model B generates decentralized networks with a larger number of components, a smaller giant component size, and a low global efficiency when compared to the other algorithms, especially compared to the centralized BA networks that have all vertices in a single component, with a medium to high global efficiency. The other three models generate networks with intermediate characteristics between B and BA models. A consequence of this finding is that the dynamics of different phenomena on these networks may differ considerably.

## 1. Introduction

The degree distribution $P(k)$, defined as the fraction of vertices in the network with degree $k$, is an important property of a complex network. In particular, the degree distribution of many real world networks [1–4] was accurately fitted by a scale-free (power-law) degree distribution

$$P(k) \sim k^{-\alpha}, \qquad (1)$$

where $\alpha$ is a scaling parameter.

A power-law degree distribution was observed, for instance, in networks of animal movements [5]. Such networks are examples of networks whose degree distribution may be either estimated using a questionnaire in which the number of contacting farm holdings is assessed or through the analysis of animal movement records. When there is a large number of farm holdings in the network and a data bank of animal movements is not available, we might assess the degree distribution using a questionnaire. From the estimated degree distribution, one may be interested in recovering approximately the real network to simulate, for instance, the potential spread of infectious diseases such as foot-and-mouth disease and bovine brucellosis, for which the network of animal movements is an important means of dissemination [6–8]. Nevertheless, the process of recovering a possible real network from the estimated degree distribution may lead to a misleading inference. The presence of a scale-free degree distribution does not guarantee that the recovered network will show the same topology as the original one. More than one method may generate a network that shows a scale-free degree distribution, and, from these different methods, networks can emerge with different structural properties, which may impact the outcomes of the simulation of dynamical phenomena on the network.

In this paper, we depart from a given degree distribution and we show how to generate networks using different algorithms and the implications in the network topology of choosing one of these algorithms to generate networks when all you have is the network's degree distribution.

A well-known method to generate a scale-free network is the preferential attachment [9, 10], in which links are added to vertices based on their degree. In this approach, a network is generated, and then the resulting power-law distribution is evaluated. We use the preferential attachment approach to generate a network with a scale-free degree distribution. Based on this distribution, networks are generated using four different algorithms (two of them proposed for the first time). To qualitatively compare these networks, we calculate some of their structural properties [11].

This paper is organized as follows. In Section 2, we discuss the calculation and properties of the chosen parameters to compare the networks. In Section 3, we describe the five algorithms used to generate the networks. In Section 4, we show the results of the calculations of the structural measures for the scale-free networks obtained. Finally, in Section 5, we discuss the implications of our findings.

## 2. Structural Properties

It is worth to mention that our objective is not to perform an extensive review of all possible metrics but just highlight some global features of different networks instead. Also, there is an unlimited set of topological measurements, and they are often correlated, implying redundancy in most of the cases [11]. We calculated the following structural properties [11]: average degree of the nearest neighbors, central point dominance [12], clustering coefficient [13], the Pearson correlation coefficient [14], and global efficiency [15]. Some of these parameters are related to local properties of the networks (average degree of the nearest neighbor and clustering coefficient), and others are related to global properties (central point dominance, global efficiency, and the Pearson correlation coefficient). All chosen parameters reflect global networks' trends and also provide a meaningful interpretation regarding the networks' dynamical properties.

### 2.1. Average Degree of the Nearest Neighbors.
The average degree of the nearest neighbors of a vertex $i$ may be calculated as

$$k_{\mathrm{nn},i} = \frac{\sum_{j=1}^{m_0} a_{ij}k_j}{k_i}, \qquad (2)$$

where $a_{ij}$ is the element $ij$ of the adjacency matrix, defined as $a_{ij} = 1$ if there is an edge between vertices $i$ and $j$ and $a_{ij} = 0$, otherwise. The average degree of the nearest neighbors checks for correlations between the degrees of different vertices. If there are no correlations, $k_{\mathrm{nn}}(k)$ is independent of $k$. When $k_{\mathrm{nn}}(k)$ is an increasing function of $k$, vertices of high degree tend to connect with vertices of high degree, and the network is classified as assortative, whereas whenever $k_{\mathrm{nn}}(k)$ is a decreasing function of $k$, vertices of high degree tend to connect with vertices of low degree, and the network is called disassortative [11].

### 2.2. Clustering Coefficient.
The clustering coefficient (CC) for undirected networks may be calculated using the following definition [1, 13]:

$$\mathrm{CC} = \frac{\sum \mathrm{CC}_i}{N}, \qquad (3)$$

where $\mathrm{CC}_i$ is defined as:

$$\mathrm{CC}_i = \frac{2}{k_i(k_i - 1)} \sum_{j,k} a_{ij}a_{ik}a_{jk}, \qquad (4)$$

and $N$ is the total number of vertices in the network.

CC reflects the network's tendency to group together nodes with common links, thus raising the number of triangles found inside the network.

### 2.3. Central Point Dominance.
The central point dominance (CPD) [12] is a measure related to the betweenness centrality of the most central vertex in a network. Its value is 0 for networks in which the betweenness centralities of all vertices are equal and 1 for the wheel or star network. The equation for the CPD is [12]

$$\mathrm{CPD} = \frac{\sum_{i=1}^{N}(B_{\max} - B_i)}{N - 1}, \qquad (5)$$

where $B_{\max}$ and $B_i$ are, respectively, the largest values of the relative betweenness centrality in the network and the relative betweenness centrality of vertex $i$. The relative betweenness centrality is the ratio between the betweenness centrality of a vertex and its maximum possible value, $(N^2 - 3N + 2)/2$, which corresponds to the betweenness of the central vertex in a star network.

CPD reflects an important network characteristic, which is the network's dependence on specific vertices to maintain its information flow. Networks with higher values of CPD rely on fewer vertices to pass their information to other vertices, while networks with lower values of CPD have their flow and pathways distributed in a more decentralized way, thus being more resilient to random vertices removal.

### 2.4. Global Efficiency.
The global efficiency (GE) is a measurement that quantifies the efficiency of the network in sending information between vertices, defined as [15]

$$\mathrm{GE} = \frac{1}{N(N-1)} \sum_{i \neq j} \frac{1}{d_{ij}}, \qquad (6)$$

where $d_{ij}$ is the shortest path length between vertices $i$ and $j$. Networks with high GE can send information much faster and to a larger number of vertices than networks with low GE.

### 2.5. Correlation Coefficient.
A detailed definition for the correlation coefficient ($r$) may be found in [14]. Basically, it is simply the Pearson correlation coefficient between the degrees at either ends of an edge, consisting of another way to determine the degree correlation, besides the average degree of the nearest neighbors.

## 3. Algorithms

To guarantee that all the networks generated follow the same degree distribution, allowing comparisons between them, we have firstly generated a network following the Barabási-Albert (BA) algorithm, and then we have applied the other algorithms to generate networks based on the degree distribution of the BA network. Due to the growth process inherent in the BA algorithm, it would be difficult or even impossible to generate a BA network from a given $P(k)$ distribution.

For the sake of completeness, we describe below all the algorithms used.

### 3.1. Barabási-Albert Model.
The algorithm of the Barabási-Albert model, described in [9], is the following.

(1) We start with a disconnected set of $m_0$ vertices.

(2) At each time step, a new vertex with $m(< m_0)$ edges is added, linking the new vertex to $m$ different vertices already in the system.

(3) When choosing the vertices to which the new vertex connects, we assume that the probability that a new vertex will be connected to vertex $i$ depends on the degree $k_i$ of vertex $i$ (preferential attachment), such that

$$p(k_i) = \frac{k_i}{\sum_{j=1}^{m_0} k_j}.$$ (7)

We have used the BA algorithm implemented in the igraph package of the *R* Statistical Software [16].

### 3.2. Molloy-Reed Model.
To generate networks using the Molloy-Reed (MR) model, we have used the following algorithm.

(1) For each vertex, we choose a degree from the distribution.

(2) At each time step, we connect randomly a pair of vertices, taking into account that the probability of selecting a vertex is directly proportional to the number of its open connections, defined as the number of remaining links [17].

(3) The previous step is repeated until there are no more open connections.

In this version of the MR algorithm, multiple edges are ignored, self-edges are not allowed, and open connections may be discarded if there is only one vertex remaining.

### 3.3. Kalisky Model.
The algorithm proposed by Kalisky et al. [17] is based on the MR model. The aim of the Kalisky algorithm is to force a hierarchy on the MR model, defining layers in the graph, as follows.

(1) A degree is assigned to each vertex.

(2) We start from the maximal degree $(K)$ vertex, which is connected to $K$ open connections. The set composed by this vertex and its neighbors is the first layer of vertices.

(3) The second layer is filled out in the same way: we connect all open connections emerging from vertices in the first layer to randomly chosen open connections.

(4) This process continues until the set of open connections is empty.

### 3.4. Model A.
In this algorithm, called hereafter as Model A (MA), the vertices are randomly sampled from a list of vertices with available links. The algorithm is as follows.

(1) We choose the vertex with the highest available degree $(h)$ in the network (in the first step, this is the vertex with the maximum degree).

(2) We connect that vertex with $h$ other vertices, randomly selected from a list with available vertices, thus exhausting the links of the chosen vertex.

(3) Steps (i) and (ii) are repeated until there are no more vertices with open connections.

### 3.5. Model B.
In this algorithm, called hereafter as Model B (MB), a vector, whose elements are the degrees of all vertices obtained from the BA degree distribution, is randomly generated. Then, the vertices are selected in sequence, following the order of the vector elements. The algorithm is as follows.

(1) We choose the vertex with the highest available degree $(h)$ in the network (in the first step, this is the vertex with the maximum degree).

(2) We connect that vertex with the first $h$ other vertices of the generated vector, thus exhausting the links of the chosen vertex.

(3) Steps (i) and (ii) are repeated until there are no more vertices with open connections.

We stress that the last two models (MA and MB) automatically avoid the generation of multiple edges and self-edges. MA generates networks with vertices connected randomly, starting the connection process with the hubs, while an interesting feature of MB is that it generates networks in which every hub is connected to the other hubs. As far as we know, these two algorithms have not been proposed before.

The computer codes used to generate the networks are available upon request. For making the codes freely available, we implemented the algorithms using the *R* Statistical Software [18], along with the Matrix package [19].

## 4. Results

Figure 1 shows the scale-free networks generated using the algorithms by Barabási-Albert (Figure 1(b)), Molloy-Reed (Figure 1(c)), Kalisky et al. (Figure 1(d)), Model A (Figure 1(e)), Model B (Figure 1(f)), and the corresponding degree distribution $P(k)$ (Figure 1(a)) based on an original network generated using the BA model with $m_0 = 1000$ vertices and adding, at each time step, a new edge between two vertices $(m = 1)$. Figure 2 shows the generated networks and the degree distribution for $m = 2$. We have used the Kamada-Kawai visualization algorithm implemented in the *R* "network" package [20].

We notice in Figure 1 that the BA network has only one component, while the other models generate networks with several components. For $m = 1$, this behavior may be observed in Figure 3. For $m = 2$, however, all models tend to generate only one giant component, with the exception

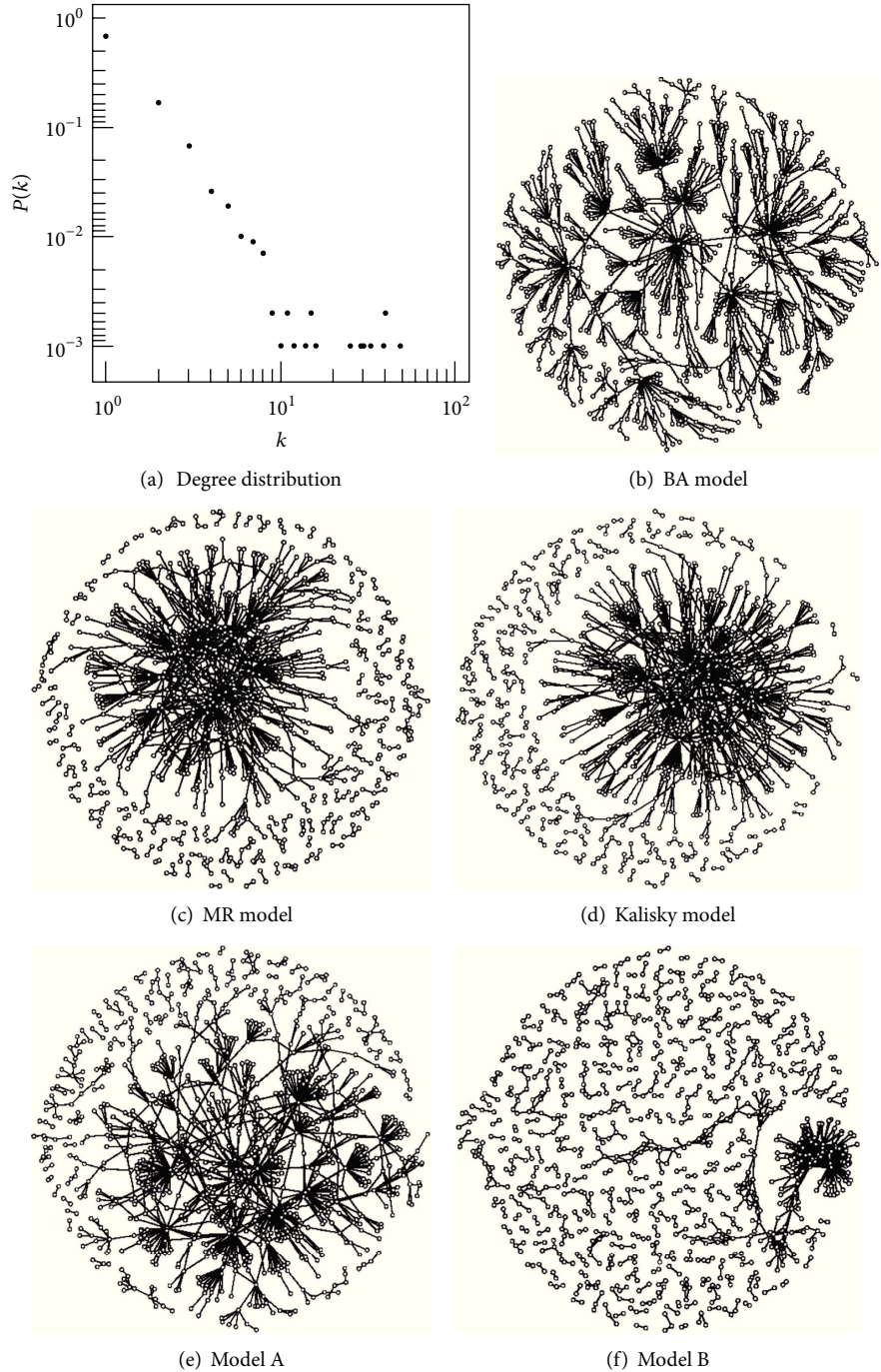

FIGURE 1: Network generated using the BA model with $m_0 = 1000$ vertices and $m = 1$ (b) and corresponding degree distribution (a), from which the other scale-free networks were derived using the following algorithms: (c) MR model, (d) Kalisky model, (e) model A, and (f) model B.

of MB, which generates a larger number of components (Figure 3).

To assess the assortativity of the different networks, we analyzed the average degree of the nearest neighbors of vertices with degree $k$, $\langle k_{nn} \rangle$, as a function of $k$ for networks with $10^3$ vertices and $m = 1$ or $m = 2$ (Figure 4). We have also analyzed networks with $10^2$, $10^3$, and $10^4$ vertices with $m = 2$ and $m = 3$, but the qualitative results were similar. As a general behavior, the algorithms used provide disassortative mixing. The exception is the network generated using MB (Figures 4(e) and 4(j)), for which an assortative mixing is observed for degrees up to a critical value (between 10 and 15), followed by a disassortative mixing onwards.

Probably due to a high level of redundancy in the giant component, for $m = 1$, the median clustering coefficient for the MB network (0.08) is higher than the values observed for the others (Figure 5(a)). Due to the topology of the BA network for $m = 1$, in which no triangles are observed, the CC is zero as expected. For all networks, both the median and the interquartile range of the CC increase for $m = 2$. The higher median CC values (around 0.12) were observed for the BA, Kalisky, and MB networks.

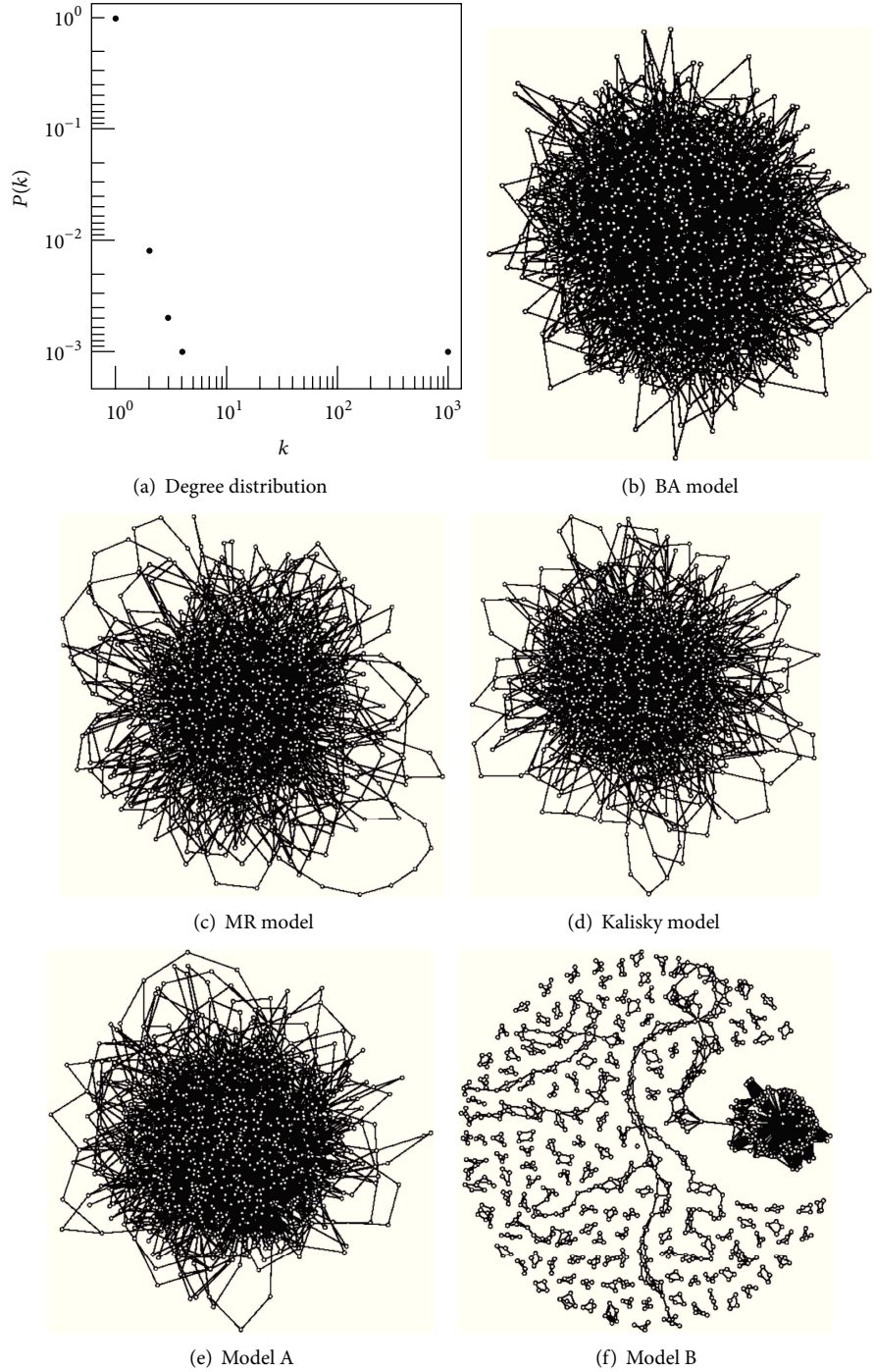

(a) Degree distribution

(b) BA model

(c) MR model

(d) Kalisky model

(e) Model A

(f) Model B

Figure 2: Network generated using the BA model with $m_0 = 1000$ vertices and $m = 2$ (b) and corresponding degree distribution (a), from which the other scale-free networks were derived using the following algorithms: (c) MR model, (d) Kalisky model, (e) model A, and (f) model B.

In Figure 5(b), we notice that the CPD is lower for the MB networks and higher for the BA networks (for $m = 1$). Moreover, the CPD values are higher for networks related to $m = 2$ when compared with the networks generated using $m = 1$. The exception is the BA network, for which the addition of edges probably reduces the betweenness centrality of the hubs, reducing the CPD value.

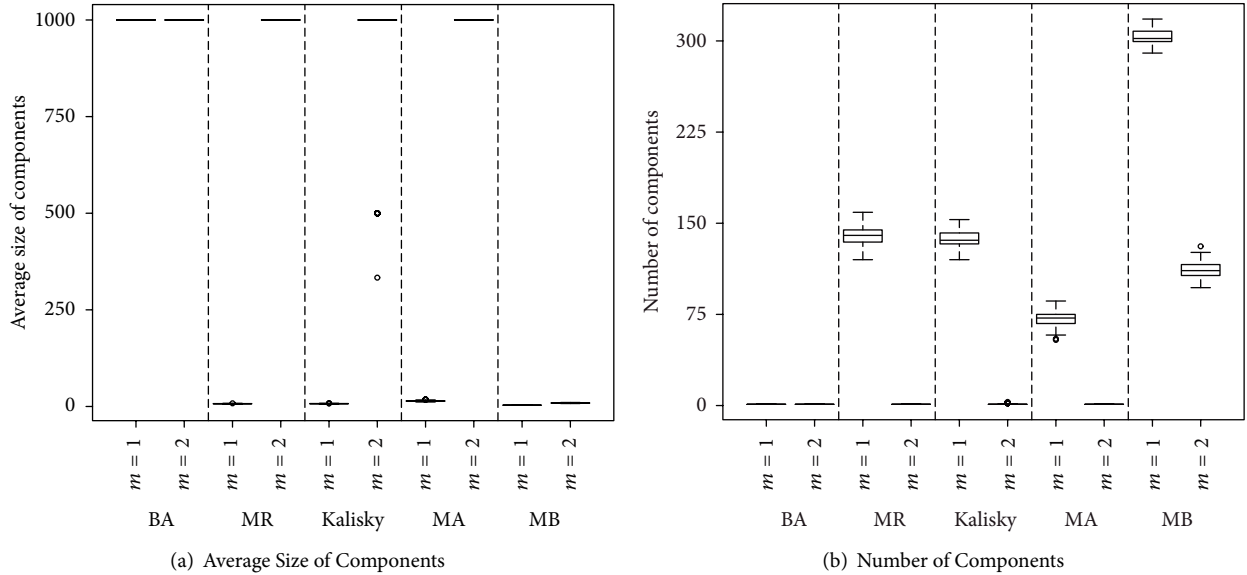

(a) Average Size of Components

(b) Number of Components

Figure 3: Boxplots for the (a) average size and (b) number of components for 100 different networks with 1000 vertices each and $m = 1$ or $m = 2$.

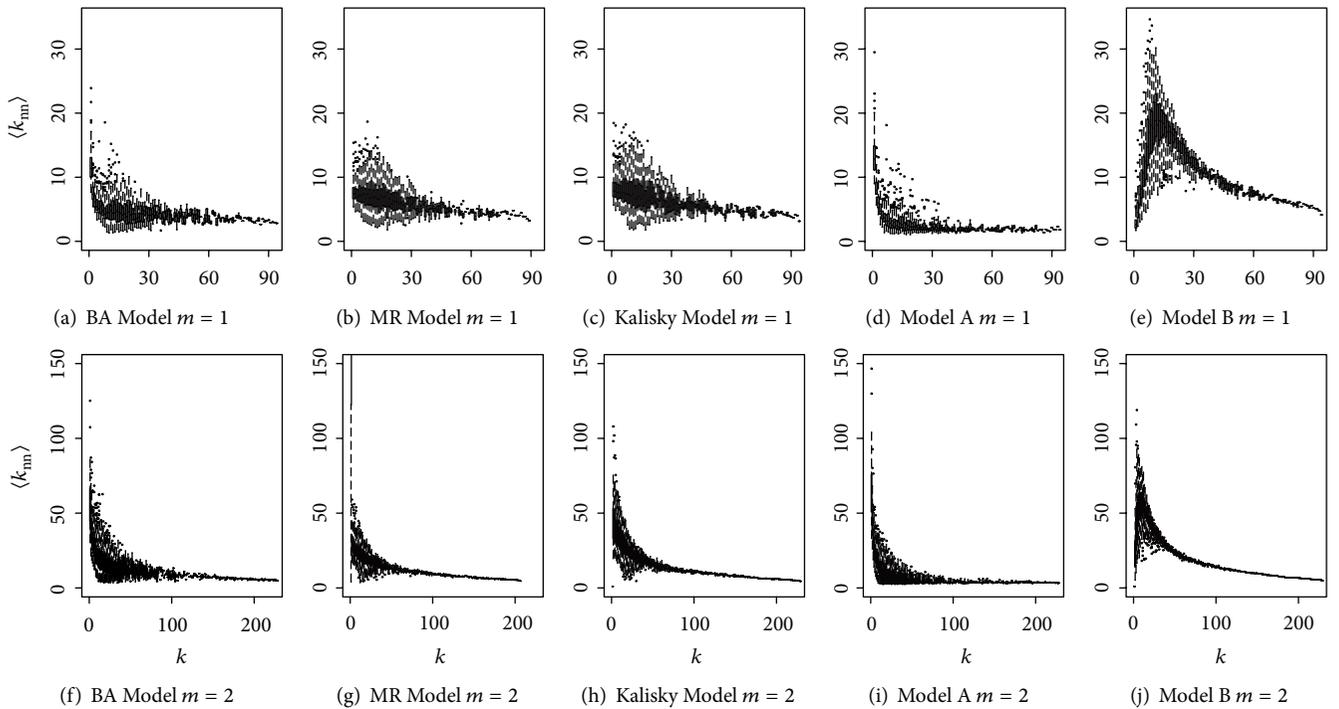

(a) BA Model $m = 1$   (b) MR Model $m = 1$   (c) Kalisky Model $m = 1$   (d) Model A $m = 1$   (e) Model B $m = 1$

(f) BA Model $m = 2$   (g) MR Model $m = 2$   (h) Kalisky Model $m = 2$   (i) Model A $m = 2$   (j) Model B $m = 2$

Figure 4: Boxplots for the average degree of the nearest neighbors $\langle k_{nn} \rangle$ as a function of the degree of a given vertex for 100 different networks with 1000 vertices and $m = 1$ ((a)–(e)) or $m = 2$ ((f)–(j)).

In Figure 5(c), we notice that the network generated using MB is clearly less efficient than the other networks due to its higher number of small components (Figure 3). On the other hand, for $m = 1$, the BA network has the highest median GE (0.08), probably because in this network there is always only one component. However, the number of components is not the only factor influencing GE, since BA network has a higher

GE for $m = 2$ than for $m = 1$, showing that the number of links also has a major impact in GE, as expected. Also, for $m = 2$, as we can see in Figure 3, with the exception of the MB network, all the networks have only one component and a similar GE (median of 0.15).

Estimates of the correlation coefficient for the different types of networks are shown in Figure 5(d). For $m = 1$, we may notice that positive values were mainly observed in the MB network. This finding is consistent with the analysis of the $\langle k_{nn} \rangle$, since negative correlation coefficients were found for the networks with a disassortative mixing pattern. For $m = 2$, negative values for the correlation coefficient were also observed for the MB network.

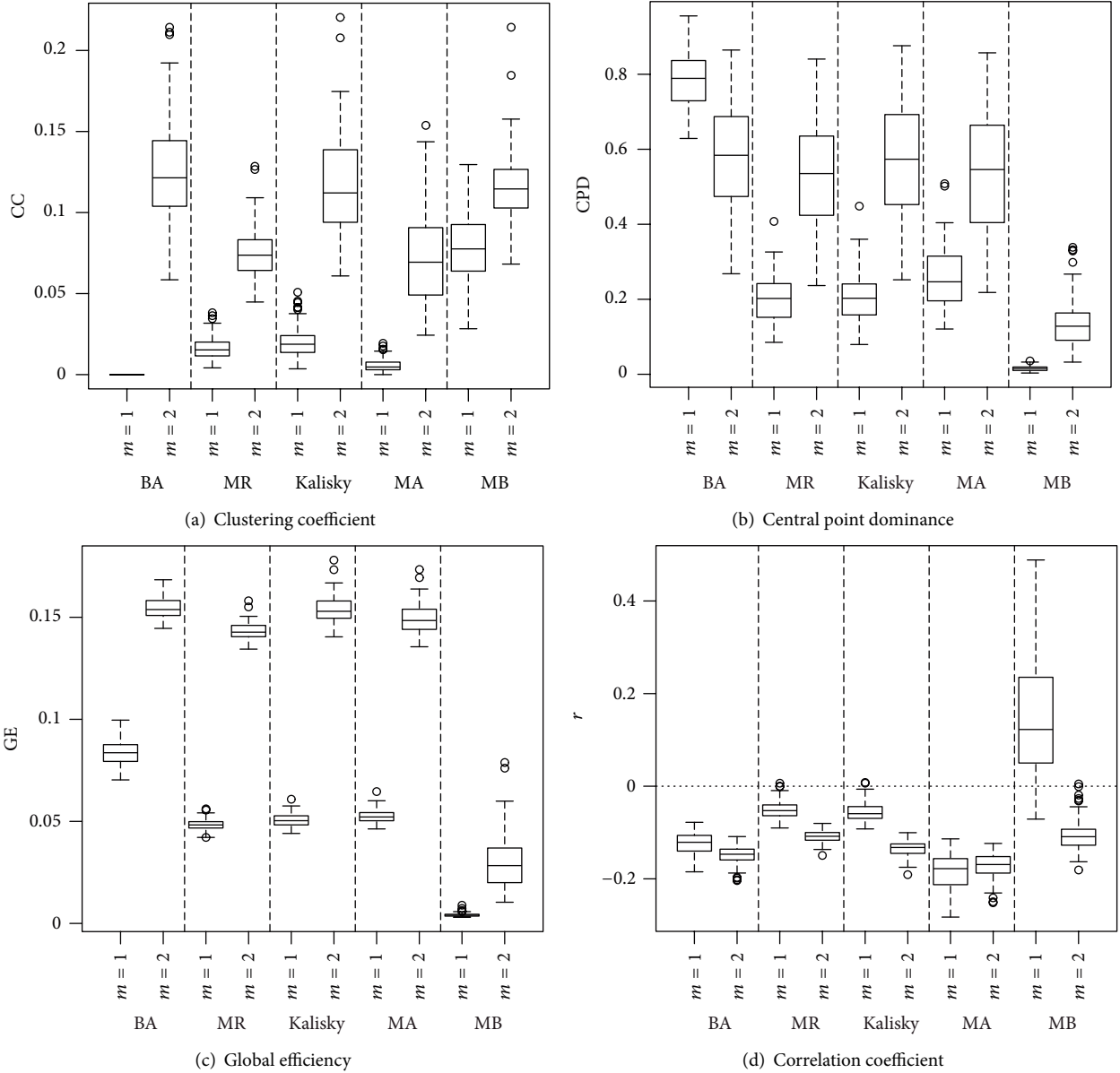

FIGURE 5: Boxplots of the following structural measures for 100 networks with 1000 vertices and $m = 1$ or $m = 2$: (a) clustering coefficient, (b) central point dominance, (c) global efficiency, and (d) correlation coefficient.

Table 1 summarizes the results of the average number of components and the average size of the giant component (in percentage of the entire network) for the five models. Comparing the models, the extreme cases are the MB and the BA networks: the MB networks show a larger number of components, a smaller giant component size, and a very low (for $m = 1$) to low (for $m = 2$) GE and CPD; while the BA networks have only one component, medium (for $m = 1$) to high (for $m = 2$) GE, and very high (for $m = 1$) to very high (for $m = 2$) CPD. The other three models analyzed generate networks with intermediate characteristics between MB and BA models but approaching the BA model when $m = 2$. In particular, for $m = 2$, regarding the average number of components, the MA networks are closer to the BA networks. MR and Kalisky networks show similar number of components and giant component size.

## 5. Concluding Remarks

We have implemented different algorithms that generate networks from a given degree distribution. As we show in the sequel of this paper [21], it is possible to generate the networks using the algorithms and then simulate the dynamics of an infectious disease on these networks. An important finding of [21] is that the simulations for the susceptible-infected-susceptible (SIS) infectious diseases models show that the disease prevalence in MB networks is lower than in the other networks, which may be related to the MB network structure, in which a large set of vertices are not connected to the main component of the network.

Table 1: Average number of components, average size (in percentage of the entire network) of the giant component (GC), and categorical classifications for GE (high: GE > 0.12; medium: 0.12 > GE > 0.05; low: 0.05 > GE > 0.01; very Low: GE < 0.01) and CPD (very high: CPD > 0.7; high: 0.7 > CPD > 0.4; medium: 0.4 > CPD > 0.2; low: 0.2 > CPD > 0.1; very low: CPD < 0.1).

| | BA | | MR | | Kalisky | | MA | | MB | |
|---|---|---|---|---|---|---|---|---|---|---|
| | $m = 1$ | $m = 2$ | $m = 1$ | $m = 2$ | $m = 1$ | $m = 2$ | $m = 1$ | $m = 2$ | $m = 1$ | $m = 2$ |
| No. components | 1.00 | 1.00 | 140.39 | 1.05 | 136.29 | 1.14 | 70.43 | 1.01 | 303.08 | 112.20 |
| GC size (%) | 100.00 | 100.00 | 67.00 | 100.00 | 69.00 | 100.00 | 80.00 | 100.00 | 17.00 | 36.00 |
| Efficiency (GE) | Medium | High | Medium | High | Medium | High | Medium | High | Very low | Low |
| Dependence (CPD) | Very High | High | Medium | High | Medium | High | Medium | High | Very low | Low |

Regarding the results observed, an aspect that calls attention is that the network generated using algorithm MB differs (by visual inspection) from the networks generated using the other models. In fact, the MB algorithm generates a network with a larger number of components and a smaller giant component size, if compared to the other algorithms, as shown in Table 1 and Figure 3.

The MB networks show lower CPD and global efficiency values, and assortative mixing for low degree values when compared to the other networks. These properties are probably a consequence of the distribution of components in the MB network, with one giant component and a large number of small components. On the other hand, for $m = 1$, the BA networks show the higher CPD and global efficiency median values, possibly reflecting the existence of only one component in these networks. For $m = 2$, a similar comment applies to all models with the exception of MB.

Based on the findings presented in this paper, we may hypothesize that, based only on the observed degree distribution $P(k)$, it may not be possible to make an accurate inference about some structural properties of the network. A consequence of this remark is that different scale-free networks (and possibly other types of networks, except lattice and similar networks) with the same degree distribution may have distinct structural properties so that the dynamics of different phenomena on these networks may differ considerably.

Different algorithms may be invented to generate networks from a given degree distribution. Provided that a network is generated, sets of vertices may be rearranged to increase or decrease the components' sizes. In this paper, we analyzed five specific algorithms, ranging from the BA model, which always generates a network with a single component, to the MB algorithm, which can generate a network with several components, and with three other intermediate cases. The effects of our findings are clearly evident, with one model (MB) giving decentralized and low efficient networks and another one (BA) giving networks much more efficient and centralized, with three cases in the middle, all of which with exactly the same degree distribution.

A word of caution is in order: when generating a scale-free network from a given degree distribution, researchers should state and, if necessary, describe clearly which algorithm was used. Otherwise, from the same $P(k)$, the simulation of dynamical phenomena can result in different outcomes depending on the algorithm used to generate the network.

Thus, for those interested in applying questionnaires to infer the network structure, based only on the degree distribution, it is possible to estimate the average degree, the degree variance and other moments of the statistical distribution, that is, properties that derive directly from the degree distribution, but it is not possible to infer the dynamical properties. If the interest is to analyse dynamical processes on the network, the degree distribution is not enough, it is necessary to have the adjacency matrix. In other words, it is necessary to know the links within the network.

## Acknowledgments

This work was partially supported by FAPESP and CNPq.

## References

[1] G. Caldarelli, *Scale-Free Networks*, Oxford University Press, Oxford, UK, 2007.

[2] L. A. N. Amaral, A. Scala, M. Barthélémy, and H. E. Stanley, "Classes of small-world networks," *Proceedings of the National Academy of Sciences of the United States of America*, vol. 97, no. 21, pp. 11149–11152, 2000.

[3] H. Jeong, "Complex scale-free networks," *Physica A*, vol. 321, pp. 226–237, 2003.

[4] A. Clauset, C. R. Shalizi, and M. E. J. Newman, "Power-law distributions in empirical data," *SIAM Review*, vol. 51, no. 4, pp. 661–703, 2009.

[5] M. Bigras-Poulin, R. A. Thompson, M. Chriel, S. Mortensen, and M. Greiner, "Network analysis of Danish cattle industry trade patterns as an evaluation of risk potential for disease spread," *Preventive Veterinary Medicine*, vol. 76, no. 1-2, pp. 11–39, 2006.

[6] J. C. Gibbens, C. E. Sharpe, J. W. Wilesmith et al., "Descriptive epidemiology of the 2001 foot-and-mouth disease epidemic in Great Britain: the first five months," *Veterinary Record*, vol. 149, no. 24, pp. 729–743, 2001.

[7] R. L. Negreiros, M. Amaku, R. A. Dias, F. Ferreira, J. C. M. Cavalléro, and J. S. F. Neto, "Spatial clustering analysis of the foot-and-mouth disease outbreaks in Mato Grosso do Sul state, Brazil—2005," *Ciência Rural*, vol. 39, no. 9, pp. 2609–2613, 2009.

[8] R. A. Dias, V. S. P. Gonçalves, V. C. F. Figueiredo et al., "Situação epidemiológica da brucelose bovina no Estado de São Paulo [Epidemiological situation of bovine brucellosis in the State of São Paulo, Brazil]," *Arquivo Brasileiro de Medicina Veterinária e Zootecnia*, vol. 61, supplement 1, pp. 118–125, 2009.

[9] R. Albert and A. L. Barabási, "Statistical mechanics of complex networks," *Reviews of Modern Physics*, vol. 74, no. 1, pp. 47–97, 2002.

[10] V. Batagelj and U. Brandes, "Efficient generation of large random networks," *Physical Review E*, vol. 71, no. 3, Article ID 036113, 5 pages, 2005.

[11] L. F. Costa, F. A. Rodrigues, G. Travieso, and P. R. Villas Boas, "Characterization of complex networks: a survey of measurements," *Advances in Physics*, vol. 56, no. 1, pp. 167–242, 2007.

[12] L. C. Freeman, "A set of measures of centrality based on betweenness," *Sociometry*, vol. 40, no. 1, pp. 35–41, 1977.

[13] D. J. Watts and S. H. Strogatz, "Collective dynamics of "small-world" networks," *Nature*, vol. 393, no. 6684, pp. 440–442, 1988.

[14] M. E. J. Newman, "Assortative mixing in networks," *Physical Review Letters*, vol. 89, no. 20, Article ID 208701, 4 pages, 2002.

[15] V. Latora and M. Marchiori, "Efficient behavior of small-world networks," *Physical Review Letters*, vol. 87, no. 19, Article ID 198701, 4 pages, 2001.

[16] G. Csardi and T. Nepusz, "The igraph software package for complex network research," *International Journal Complex Systems*, no. 1695, pp. 1–9, 2006.

[17] T. Kalisky, R. Cohen, D. ben-Avraham, and S. Havlin, "Tomography and stability of complex networks," *Lecture Notes in Physics*, vol. 650, pp. 3–34, 2004.

[18] R Core Team, *R: A Language and Environment for Statistical Computing*, R Foundation for Statistical Computing, Vienna, Austria, 2012, http://www.R-project.org/.

[19] D. Bates and M. Maechler, "Matrix: sparse and dense matrix classes and methods," http://CRAN.Rproject.org/package=Matrix.

[20] C. T. Butts, "Network: a package for managing relational data in R," *Journal of Statistical Software*, vol. 24, no. 2, pp. 1–36, 2008.

[21] R. Ossada, J. H. H. Grisi-Filho, M. Amaku, and F. Ferreira, "Modeling the dynamics of infectious diseases in scale-free networks with the same degree distribution," 2013, submitted.